\documentclass{PoS}

\title{Decomposing Feynman rules}

\ShortTitle{Decomposing Feynman rules}

\author{Francis Brown\thanks{FB is partially supported by ERC grant no.\ 257638.}\\
        Humboldt U.\ and IHES, 91440 Bures-sur-Yvette, France\\
        E-mail: \email{brown@math.jussieu.fr}}

\author{\speaker{Dirk Kreimer}%
         \thanks{DK thanks the Alexander von Humboldt foundation and the BMBF for support through an Alexander von Humboldt Professorship.}\\
        Inst.\ of Physics and Inst. of Math.\, Humboldt Univ., 10099 Berlin, Germany\\
        E-mail: \email{kreimer@physik.hu-berlin.de}}

\def\One{\mathbb{I}}
\newtheorem{defi}{Definition}
\newtheorem{thm}{Theorem}

\abstract{We exhibit in the simple example of the Dunce's cap Feynman graph the structure of parametric renormalization on the level of integrands,
and exhibit also  the decomposition $\Phi^R=\Phi_{\mathrm{fin}}^{-1}(\Theta_0)\star \Phi^R_{\mathrm{\makebox{1-s}}}(S/S_0)\star\Phi_{\mathrm{fin}}(\Theta)$
into angles and scales of renormalized Feynman rules.}

\FullConference{{\it Loops and Legs in Quantum Field Theory} - 11th DESY Workshop on Elementary Particle Physics \\
		 April 15-20, 2012\\
		 Wernigerode, Germany}

\usepackage{epsf}
\def\dunce{\;\raisebox{-12mm}{\epsfysize=36mm\epsfbox{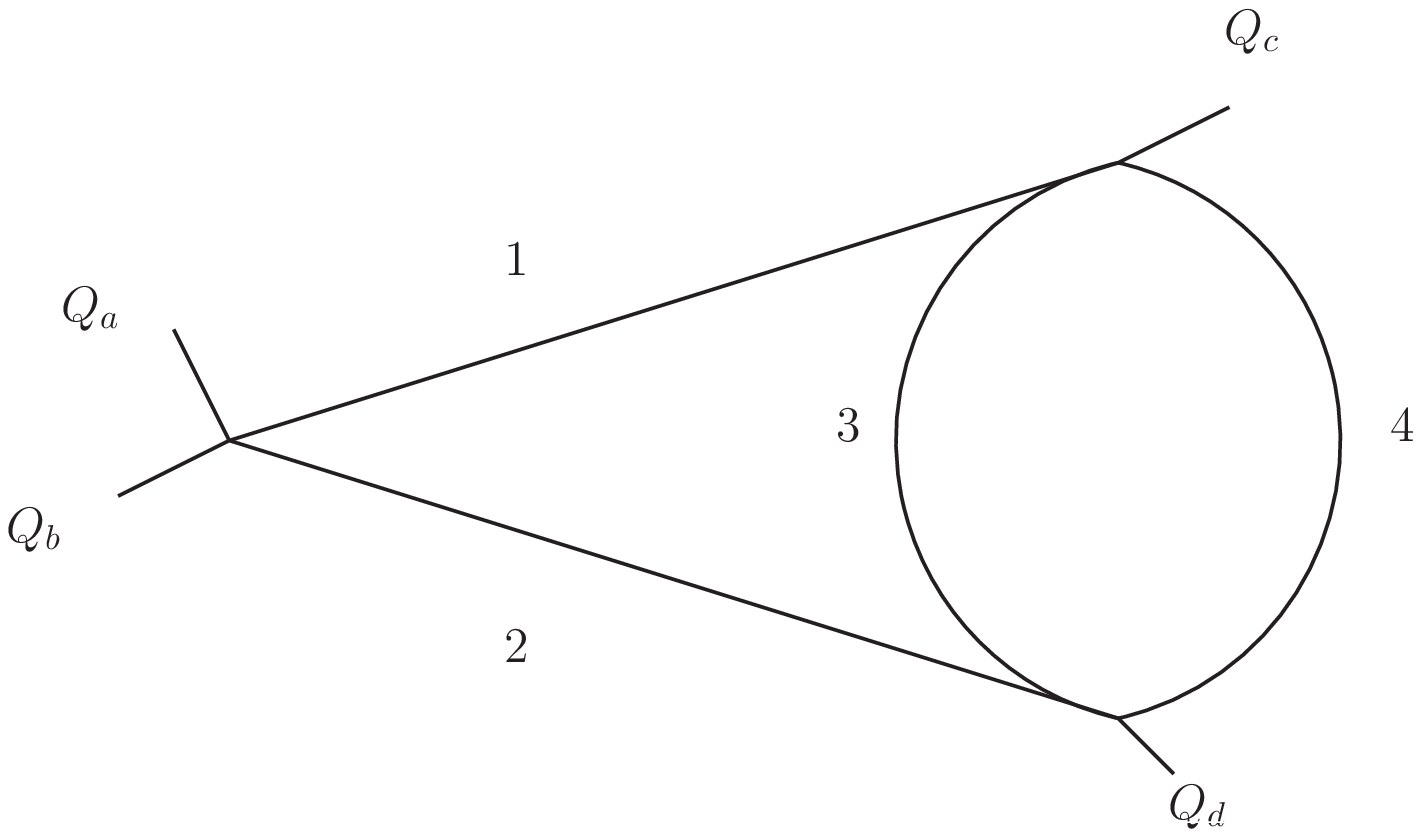}}\;}

\begin{document}

\section{Introduction}
This short paper reviews the results of \cite{BrKr}, has some comments  in the view of \cite{KrSavS}, and exhibits an example.
After the  presentation at LL2012, in which there was neither space nor time to exhibit a projectively renormalized Feynman integral with subdivergences,
the example below will answer Kostja Chetyrkin's question for such an example. We stress though that the results in \cite{BrKr} are proven by mathematical standards of rigour
for any Feynman graph in a scalar field theory, and that the results of \cite{KrSavS} further generalize them to gauge theories.  
\section{Algebra of graphs}
We consider connected scalar graphs as generators of a free commutative $\mathbb{Q}$-algebra of graphs $H$. 
We write $\One$ for the unit represented by the empty set, with disjoint union of graphs furnishing the product.
Graphs $\Gamma$ have external edges $E^\Gamma_E$, internal edges $E^\Gamma_I$ and vertices $V^\Gamma$, as usual.
For a graph $\Gamma$, we define its weight $\omega^\Gamma=4-|E^\Gamma_E|$. We allow vertices to be three- or four-valent.
Note that these weights reproduce the superficial degree of divergence  of scalar $\phi^4$ graphs and of gauge theory graphs in four dimensions simultaneously
(for graphs with external bosons only).

We say that a graph is positive-valued if $\omega^\Gamma\geq 0$. 
A graph has $\iota$-valuation if $\omega^\Gamma\geq\iota$.
The valuation of a graph is invariant under shrinking edges.

Valuations give rise to Hopf algebras. Here we use the Hopf algebra originating from positive valuation:
\begin{equation}
\Delta_0\Gamma=\Gamma\otimes\One+\One\otimes\Gamma+\sum_{\emptyset\not=\gamma=\prod\gamma_i,\omega^{\gamma_i}\geq 0}\gamma\otimes \Gamma/\gamma,
\end{equation} 
is a co-product, for a connected commutative Hopf algebra with unit $\One$ and the span of all non-trivial graphs as augmentation ideal \cite{Kr}.
The 2-valuation gives a different Hopf algebra with co-product
\begin{equation}
\Delta_2\Gamma=\Gamma\otimes\One+\One\otimes\Gamma+\sum_{\emptyset\not=\gamma=\prod\gamma_i,\omega^{\gamma_i}\geq 2}\gamma\otimes \Gamma/\gamma,
\end{equation} 
which is useful in treating the quadratic divergences of self-energy graphs \cite{BrKr}.

\begin{defi}\label{forest}
A forest $f$,  $f:=\{\gamma_i\}$ is a subset of proper positive valued 1PI sub-graphs $\gamma_i\subset\Gamma$ such that any two elements $\gamma_i,\gamma_j$
of $f$ fulfil: $\gamma_i\cap\gamma_j=\emptyset$, or $\gamma_i\subset\gamma_j$ or $\gamma_j\subset\gamma_i$. 
\end{defi}

\begin{defi}
It is maximal, iff $\Gamma/f$ contains no positive valued proper sub-graph. It is complete, if it contains all positive valued proper sub-graphs of all its elements.
\end{defi}

$|f|$ is the number of elements of $f$ and $\mathcal{F}^\Gamma$ the set of all forests of $\Gamma$.

The antipode corresponding to the co-product $\Delta_0$ can be written as
\begin{equation}
S(\Gamma)=-\sum_{f\in\mathcal{F}^\Gamma}(-1)^{|f|}f\times \Gamma/f,
\end{equation}
where the sum includes the empty set.

If the number of external edges of a sub-graph $\gamma$ is greater than two, $|E^\gamma_E|>2$, $\gamma$ shrinks to a
vertex in $\Gamma/\gamma$. If it equals two, the two external edges are identified to a single edge in $\Gamma/\gamma$ \footnote{If we were to have massive particles,
a more elaborate notation is needed \cite{BrKr}.}.

\section{Feynman rules}
\subsection{The unrenormalized integrand}
We are considering Feynman rules in parametric space.
We denote the first Symanzik polynomial by $\psi_\Gamma$, the second one by $\phi_\Gamma$. Edge variables are denoted as $A_e$ for edges $e$.
(We have $A_e\in\mathbb{R}_+$.)

 A  parametric integrand $I_\Gamma$ and a hypercube $\sigma_\Gamma=\{A_e:A_e>0\}$  written as in 
\[
\Phi(\Gamma) = \int_{\sigma_\Gamma}  I_\Gamma \prod_e d\!A_e,
\]
for now means just that: a pair of the hypercube and the form. It is to be regarded as an honest integral only when the integrand is replaced by its suitably renormalized form
as defined below,
typically indicated by a superscript ${}^R$, 
so that the integral actually exists. We often simply write $\int$ for $\int_{\sigma_\Gamma}$.

In Schwinger parametric form, the unrenormalized  Feynman amplitude is (omitting trivial overall factors of powers of $\pi$ and such)
\begin{equation}\label{param}
\Phi(\Gamma) = \int \underbrace{\frac{e^{-\frac{\phi_\Gamma}{\psi_\Gamma}}}{\psi_\Gamma^2}}_{I_\Gamma}\prod_e d\! A_e.
\end{equation}
\subsection{The renormalized integrand}
We can render this pairing of $\sigma_\Gamma$ with $I_\Gamma$ integrable by a suitable sum over forests. We define
\[
I^R_\Gamma:=\sum_{f\in\mathcal{F}_\Gamma}(-1)^{|f|} I_{\Gamma/f} I_f^0,
\]
where for $f=\bigcup_i\gamma_i$, $I_f=\prod_i I_{\gamma_i}$ and the superscript ${}^0$ indicates that kinematic variables are specified according to renormalization conditions. The formula for $I_\Gamma^R$ is correct as long as all sub-graphs are overall log-divergent, the necessary correction terms in the general case are given in \cite{BrKr}.

\subsection{The renormalized integral}
Accompanying this integrand is the renormalized result which can be written projectively:
\[
\Phi^R(\Gamma):=\int_{\mathbb{P}^{|E^\Gamma_I|}(\mathbb{R}_+)}\sum_{f\in\mathcal{F}_\Gamma}(-1)^{|f|} \frac{\ln{\frac{\phi_{\Gamma/f}\psi_f+\phi^0_f\psi_{\Gamma/f}}{\phi_{\Gamma/f}^0\psi_f+\phi^0_f\psi_{\Gamma/f}}}}{\psi^2_{\Gamma/f}\psi^2_f}\Omega_\Gamma,
\]
for notation see \cite{BrKr} or \cite{KrSavS}.

Note that this is a well-defined integral obtained from the use of the forest formula.
It is obtained without using an intermediate regulator. It is well-suited to analyse the mathematical structure
of perturbative contributions to Green functions. Also, combining this approach
with \cite{KrSavS}, it furnishes a reference point against which to check in a situation where intermediate regulators are spoiling the symmetries of the theory.
\section{Scales and angles}
Feynman graphs have their external edges labelled by momenta, and internal edges labelled by masses. 

Renormalized Feynman rules above are therefore functions of scalar products $Q_i\cdot Q_j$ and mass-squares $m_e^2$.
Equivalently, upon defining a positive definite  linear combination $S$ of such variables, we can write them as function of such a scale $S$, and angles
$\Theta_{ij}:=Q_i\cdot Q_j/S$, $\Theta_e:=m_e^2/S$. We use $S^0,\Theta^0_{ij},\Theta^0_e$ to specify the renormalization point. A graph which furnishes only a single scalar
product $Q\cdot Q$ as a scale is denoted a 1-scale graph.

Following \cite{BrKr}, we have the decomposition
\begin{thm}
\[
\Phi^R(S/S^0,\{\Theta,\Theta^0\})=\Phi_{\mathrm{fin}}^{-1}(\{\Theta^0\})\star\Phi_{\mathrm{\makebox{1-s}}}^R(S/S^0)\star\Phi_{\mathrm{fin}}(\{\Theta\}).
\]
\end{thm}
Here, the angle-dependent Feynman rules $\Phi_{\mathrm{fin}}$ are computed by eliminating short-distance singularities through the comparison, via the Hopf algebra,
with 1-scale graphs evaluated at the same scale as the initial graphs, while the 1-scale Feynman rules $\Phi_{\mathrm{\makebox{1-s}}}^R(S/S_0)$ eliminate short-distance singularities by renormalizing 1-scale graphs at a reference scale $S_0$. It is the purpose of this note to make this decomposition -treated at length in \cite{BrKr}-
accessible to the reader.

\section{The example: Dunce's cap}
We consider the graph $\Gamma$ (with sub-graph $\gamma$ spanned by edges $3,4$):
\[
\Gamma=\dunce
\]
We find the following graph polynomials:
\[
\phi_\Gamma=(Q_a+Q_b)^2(A_3+A_4)A_1A_2+Q_c^2 A_1A_3A_4+Q_d^2 A_2A_3A_4,
\]
and
\[
\psi_\Gamma=(A_1+A_2)(A_3+A_4)+A_3A_4
\]
for $\Gamma$.
For the sub-graph $\gamma$,
$
\phi_{\gamma}=Q_d^2A_3A_4,
$
and
$
\psi_{\gamma}=A_3+A_4.
$
For the co-graph $\Gamma/\gamma$, we have
$
\phi_{\Gamma/\gamma}=(Q_a+Q_b)^2A_1A_2,
$
and
$
\psi_{\Gamma/\gamma}=A_1+A_2.
$

We also define (to fix symmetric renormalization conditions)
\[
\phi_\Gamma^0=\mu^2\left(\frac{4}{3}(A_3+A_4)A_1A_2+ A_1A_3A_4+ A_2A_3A_4\right),
\]
\[
\phi_{\Gamma/\gamma}^0=\frac{4}{3}\mu^2A_1A_2,
\]
\[
\phi_{\gamma}^0=\frac{4}{3}\mu^2A_3A_4,
\]
and $\tilde\phi_X=\phi_X+\psi_X\sum_{e\in X}m_e^2 A_e$, 
$\tilde\phi_X^0=\phi_X^0+\psi_X\sum_{e\in X}m_e^2 A_e$, for all $X\in \Gamma,\Gamma/\gamma,\gamma$.

We have the standard projective form $\Omega_\Gamma$,
\[
\Omega_\Gamma=A_1 dA_2\wedge dA_3\wedge dA_4-A_2dA_1\wedge dA_3\wedge dA_4+A_3dA_1\wedge dA_2\wedge dA_4-A_4dA_1\wedge dA_2\wedge dA_3.
\]
As there is only a single divergent subgraph, the sum over forests gives the renormalized integrand as
\[
I^R_\Gamma:=\frac{\ln{\frac{\tilde\phi_\Gamma}{\tilde \phi^0_\Gamma}}}{\psi_\Gamma^2}-\frac{\ln{\frac{\tilde\phi_{\Gamma/\gamma}\psi_\gamma+
\tilde\phi_{\gamma}^0\psi_{\Gamma/\gamma}}{\tilde\phi_{\Gamma/\gamma}^0\psi_\gamma+
\tilde\phi_{\gamma}^0\psi_{\Gamma/\gamma}}}}{\psi_{\Gamma/\gamma}^2\psi_\gamma^2}.
\]
Then,
\begin{equation}
\Phi^R(\Gamma):=\int_{\mathbb{P}^3(\mathbb{R}_+)}I^R_\Gamma \Omega_\Gamma.\label{finres}
\end{equation}
is well-defined and fulfils symmetric renormalization conditions: it vanishes at $Q_i\cdot Q_j=\frac{4\delta_{ij}-1}{3}\mu^2$, $ i,j\in a,b,c,d$.

We can now turn to affine variables by setting any chosen edge variable to $1$, and carry out the remaining affine integration,
confirming the `Cheng-Wu theorem'  for the renormalized integrand.

The naked eye then confirms existence of these integrals, using nothing more than elementary properties of the logarithm function and the remainder properties of the graph polynomials:
\[
\psi_\Gamma=\psi_{\Gamma/\gamma}\psi_\gamma+R^\Gamma_\gamma,\,\,\tilde\phi_\Gamma=\tilde \phi_{\Gamma/\gamma}\psi\gamma+\tilde R^\Gamma_\gamma.
\]
Here, the remainders $R^\Gamma_\gamma,\tilde R^\Gamma_\gamma$ are of higher degree in the $\gamma$-variables than $\psi_\gamma$ itself.
This indeed suffices to prove renormalizability and the renormalization group equation by methods of algebraic geometry and the use of the Hopf algebra $H$, see  \cite{BrKr}.

Note that were we to regularize dimensionally before renormalization, we could not avoid to consider sector decompositions before making sense out of the regularized integrand.

Let us now turn to the decomposition into scales and angles.
We switch to variables $S=2(Q_a^2+Q_b^2+Q_c^2+Q_a\cdot Q_b+Q_b\cdot Q_c+Q_c\cdot Q_a)$, $S^0=4\mu^2$,
$\Theta_{i,j}=Q_i\cdot Q_j/S$, $\Theta_e=m_e^2/S$, $\Theta_{i,j}^0=(4\delta_{ij}-1)/12$, $\Theta_e^0=m_e^2/(4\mu^2)$.  

Furthermore, the co-product gives us ($\Delta_0^2:=(\Delta_0\otimes\mathrm{id})\circ\Delta_0$)
\[
\Delta^2_0(\Gamma)=\Gamma\otimes \One\otimes\One+\One\otimes\Gamma\otimes\One+\One\otimes\One\otimes\Gamma+\gamma\otimes\Gamma/\gamma\otimes\One
+\gamma\otimes\One\otimes\Gamma/\gamma+\One\otimes\gamma\otimes\Gamma/\gamma.
\] 
This has to be evaluated using
$\Phi_{\mathrm{fin}}^{-1}(\{\Theta_0\})\otimes \Phi_{\mathrm{\makebox{1-s}}}^R(S/S_0)\otimes \Phi_{\mathrm{fin}}(\{\Theta\})$, with the set of variables above.
We find
\begin{eqnarray}\label{exa1}
\Phi^R(\Gamma) & = & \Phi_{\mathrm{fin}}^{-1}(\Theta_0)(\Gamma)+\Phi_{\mathrm{\makebox{1-s}}}^R(\Gamma)+\Phi_{\mathrm{fin}}(\Theta)(\Gamma)\nonumber\\
 & & +\Phi_{\mathrm{fin}}^{-1}(\Theta_0)(\gamma)\Phi^R_{\mathrm{\makebox{1-s}}}(\Gamma/\gamma) +\Phi_{\mathrm{fin}}^{-1}(\Theta_0)(\gamma)\Phi_{\mathrm{fin}}(\Theta)(\Gamma/\gamma)+\Phi^R_{\mathrm{\makebox{1-s}}}(\gamma)
\Phi_{\mathrm{fin}}(\Theta)(\Gamma/\gamma)\nonumber.
\end{eqnarray}
Define (so that $\tilde\phi_\Gamma=SX(\Theta),\,\tilde\phi_\Gamma^0=S^0X^0(\Theta)$):
\[
X=X(\Theta):=(\Theta_{aa}+2\Theta_{ab}+\Theta_{bb})A_1A_2(A_3+A_4)+\Theta_{cc}A_1A_3A_4+\Theta_{dd}A_2A_3A_4+\psi_\Gamma\sum_{i=1}^4 \Theta_i A_i,
\]
\[
X^0=X^0(\Theta^0):=\frac{1}{3}A_1A_2(A_3+A_4)+\frac{1}{4}\left(A_1A_3A_4+A_2A_3A_4\right)+\psi_\Gamma \sum_{i=1}^4 \Theta_i^0 A_i,
\]
and the 1-scale graph polynomial (which is $\psi_{\Gamma^\bullet}$, where $\Gamma^\bullet$ is the graph obtained by identifying the left and lower right vertices in the Dunce's cap above)
\[
Y:=A_2(A_1A_3+A_1A_4+A_3A_4).
\]
We then have for the six contributions in Eq.(\ref{exa1}) ($\mathbb{P}^3\equiv\mathbb{P}^3(\mathbb{R}_+),
\mathbb{P}^1\equiv\mathbb{P}^1(\mathbb{R}_+)$):
\begin{eqnarray*}
 & & \Phi_{\mathrm{fin}}(\Gamma)(\Theta^0)^{-1}  =  -\int_{\mathbb{P}^3}\left(\frac{\ln\frac{X^0(\Theta)}{Y}}{\psi_\Gamma^2}-\frac{\ln\frac{(\frac{1}{3}A_1A_2+(\Theta_1^0A_1+\Theta_2^0A_2)(A_1+A_2))(A_3+A_4)+A_3A_4(A_1+A_2)}{A_1A_2(A_3+A_4)+A_3A_4(A_1+A_2)}}{\psi_{\Gamma/\gamma}^2\psi_\gamma^2}\right) \Omega_\Gamma\\
 & & +
\int_{\mathbb{P}^1}\frac{\ln\frac{A_3A_4+(\Theta_3^0A_3+\Theta_4^0A_4)(A_3+A_4)}{A_3A_4}}{\psi_\gamma^2} \Omega_{\gamma}
\int_{\mathbb{P}^1}\frac{\ln\frac{A_1A_2+(\Theta_1^0A_1+\Theta_2^0A_2)(A_1+A_2)}{A_1A_2}}{\psi_{\Gamma/\gamma}^2} \Omega_{\Gamma/\gamma},\\
 & & \Phi^R_{\mathrm{\makebox{1-s}}}(\Gamma)(S/S^0)  =  \int_{\mathbb{P}^3}\left(\frac{\ln\frac{S}{S^0}}{\psi_\Gamma^2}-\frac{\ln\frac{\frac{S}{S^0}A_1A_2(A_3+A_4)+A_3A_4(A_1+A_2)}{A_1A_2(A_3+A_4)+A_3A_4(A_1+A_2)}}{\psi_{\Gamma/\gamma}^2\psi_\gamma^2}\right) \Omega_\Gamma,\\
 & & \Phi_{\mathrm{fin}}(\Gamma)(\Theta)  =  \int_{\mathbb{P}^3}\left(\frac{\ln\frac{X(\Theta)}{Y}}{\psi_\Gamma^2}
-\frac{\ln\frac{((\Theta_{aa}+2\Theta_{ab}+\Theta_{bb})A_1A_2+(\Theta_1 A_1+\Theta_2 A_2)(A_1+A_2))(A_3+A_4)+A_3A_4(A_1+A_2)}{A_1A_2(A_3+A_4)+A_3A_4(A_1+A_2)}}{\psi_{\Gamma/\gamma}^2\psi_\gamma^2}\right) \Omega_\Gamma,\\
 & & \Phi_{\mathrm{fin}}^{-1}(\Theta_0)(\gamma)\Phi^R_{\mathrm{\makebox{1-s}}}(\Gamma/\gamma)  = 
-\int_{\mathbb{P}^1}\frac{\ln\frac{A_3A_4+(\Theta_3^0A_3+\Theta_4^0A_4)(A_3+A_4)}{A_3A_4}}{\psi_{\gamma}^2} \Omega_{\gamma}
\int_{\mathbb{P}^1}\frac{\ln\frac{S}{S^0}}{\psi_{\Gamma/\gamma}^2} \Omega_{\Gamma/\gamma},\\
 & & \Phi_{\mathrm{fin}}^{-1}(\Theta_0)(\gamma)\Phi_{\mathrm{fin}}(\Theta)(\Gamma/\gamma)
  = 
-\int_{\mathbb{P}^1}\frac{\ln\frac{A_3A_4+(\Theta_3^0A_3+\Theta_4^0A_4)(A_3+A_4)}{A_3A_4}}{\psi_{\gamma}^2} \Omega_{\gamma}\times\\
 & & \times 
\int_{\mathbb{P}^1}\frac{\ln \frac{(\Theta_{aa}+2\Theta_{ab}+\Theta_{bb})A_1A_2+(\Theta_1 A_1+\Theta_2 A_2)(A_1+A_2)}{A_1A_2}}{\psi_{\Gamma/\gamma}^2} \Omega_{\Gamma/\gamma},\\
 & & \Phi^R_{\mathrm{\makebox{1-s}}}(\gamma)
\Phi_{\mathrm{fin}}(\Theta)(\Gamma/\gamma)  = 
\int_{\mathbb{P}^1}\frac{\ln \frac{S}{S^0}}{\psi_{\gamma}^2} \Omega_{\gamma}
\int_{\mathbb{P}^1}\frac{\ln\frac{(\Theta_{aa}+2\Theta_{ab}+\Theta_{bb})A_1A_2+(\Theta_1A_1+\Theta_2A_2)(A_1+A_2)}{A_1A_2}}{\psi_{\Gamma/\gamma}^2} \Omega_{\Gamma/\gamma},
\end{eqnarray*}
which indeed adds up to (\ref{finres}).
This finishes our example. Differentiating wrt $S/S_0$ confirms periods to emerge in renormalization group functions \cite{Brown}, whilst the whole set-up allows for an algorithm described in \cite{Bogner}.

\bibliographystyle{plain}
\renewcommand\refname{References}

\end{document}